\documentclass{article}
\usepackage[utf8]{inputenc}
\usepackage{graphics}
\usepackage{graphicx}
\usepackage[margin=2cm]{geometry}
\usepackage{caption}
\usepackage{subcaption}
\usepackage{color}
\usepackage{natbib}
\usepackage{amsmath}

\newcommand{\nd}{\noindent}

\title{Evidence of Nonlinear Signatures in Solar Wind Proton Density at the L1 Lagrange Point}
\author{\small{ Dario Javier Zamora$^{1}$\thanks{E-mail: djzamora@conicet.gov.ar}, Facundo Abaca$^{1}$, Bruno S. Zossi$^{1}$, Ana G. Elias$^{1}$}, \\
\small{$^1$ Instituto de Fisica del Noroeste Argentino, CONICET and Universidad Nacional de Tucum\'an}\\
\small{ Av. Independencia 1800, Tucuman, CP 4000, Argentina}}

\date{\today}

\begin{document}

\maketitle

\begin{abstract}
The solar wind is a medium characterized by strong turbulence and significant field fluctuations on various scales. Recent observations have revealed that magnetic turbulence exhibits a self-similar behavior. Similarly, high-resolution measurements of the proton density have shown comparable characteristics, prompting several studies into the multifractal properties of these density fluctuations.

In this work, we show that low-resolution observations of the solar wind proton density over time, recorded by various spacecraft at Lagrange point L1, also exhibit non-linear and multifractal structures. The novelty of our study lies in the fact that this is the first systematic analysis of solar wind proton density using low-resolution (hourly) data collected by multiple spacecraft at the L1 Lagrange point over a span of 17 years.

Furthermore, we interpret our results within the framework of non-extensive statistical mechanics, which appears to be consistent with the observed nonlinear behavior. Based on the data, we successfully validate the q-triplet predicted by non-extensive statistical theory. To the best of our knowledge, this represents the most rigorous and systematic validation to date of the q-triplet in the solar wind.
\end{abstract}

\section{Introduction}

The solar wind refers to a low-density, high-speed stream of charged particles that emanates from the Sun and permeates the heliosphere. With the advent of space exploration, numerous missions have been dedicated to measuring the parameters and fields of the solar wind. These missions offer a unique opportunity to gain valuable insight into the nature and behavior of this phenomenon. 

Many processes occurring in the solar wind are inherently non-linear. To study these processes, it is essential to consider the temporal variations in the characteristics of the solar wind. While most traditional methods of classical physics are primarily suited for stationary or quasi-stationary phenomena, the analysis of dynamic regimes, fluctuations, and self-similar scaling requires the application of nonlinear dynamics. In this context, the development of methods based on fractal geometry to describe the temporal behavior of the solar wind is of particular interest. 

Turbulence is a fundamental feature of the solar wind and is commonly observed in both neutral fluid and plasma flows. It acts as a mechanism for transferring energy from large scales, where it is initially injected, to smaller scales. At these microscopic levels, dissipative and dispersive processes convert the transferred energy into other forms, such as heat or particle acceleration. The resulting fluctuations exhibit power law scaling, a behavior that is derived from the inherent scale invariance and self-similarity of the system. This power-law spectrum has been consistently observed in the magnetic-field fluctuations of the solar wind since the early days of space exploration, making it a well-established characteristic \citep{Coleman1968,Bruno2013}.
Numerous studies have analyzed fluctuations in the heliospheric magnetic field strength $B$ (see, for example, \citep{Burlaga1991}). These analyses have led to the identification of three key implications: (i) fat-tailed (non-Gaussian) distributions associated with energetic particle events, (ii) slow relaxation processes indicative of long-term memory effects and, (iii) multifractal structure in the time series. 

i- The non-uniform nature of energy transfer in turbulent systems causes energy to concentrate in localized spatial regions, leading to the emergence of highly energetic fluctuations. As a result, the tails of the probability distribution functions (PDFs) become populated by these energetic particles, giving rise to the so-called long or fat-tailed distributions. Such fat-tailed behavior has also been observed in the velocity distribution profiles of electrons in the solar wind \citep{Maksimovic1997,Shan2017}.
Electron distribution functions in the solar wind consistently exhibit three distinct components: a thermal core and a suprathermal halo, both present at all pitch angles, and a sharply field-aligned ‘strahl’ component, typically directed antisunward \citep{Stverak2009}. Although Coulomb collisions can account for the relative isotropy of the core population, the origin of the halo, particularly its sunward-directed portion, remains poorly understood \citep{Maksimovic2005}. Furthermore, non-Gaussian distributions have been reported in studies of electron temperature anisotropy in the solar wind \citep{Stverak2008}.
Higher statistical moments become particularly relevant when the distribution exhibits heavy tails, especially in regimes where fluctuations are comparable to or exceed the mean field. This kind of distribution was also observed in the distributions of heliospheric magnetic field strength fluctuations \citep{Burlaga2004, Burlaga2004a, Burlaga2009, Burlaga2005a}. 

ii- There is growing evidence that the transition to a quasi-equilibrium, non-Gaussian state in the solar wind involves inherently slow relaxation processes \citep{Servidio2014,Verscharen2019}. As suggested by both hydrodynamic theory and recent magnetohydrodynamic (MHD) numerical simulations, these relaxation processes can occur during the turbulent cascade and manifest as localized patches exhibiting equilibrium-like configurations. The coupling of processes across multiple scales plays a crucial role in shaping the global dynamics and thermodynamics of the solar wind. In particular, the presence of slow relaxation processes is often associated with the emergence of fat-tailed distributions \citep{Zamora2022}. 

iii- The solar wind is a highly turbulent medium, exhibiting strong field fluctuations across a broad range of scales. These include an inertial range where a turbulent cascade is believed to be active. Notably, the solar wind cascade displays intermittency, although the degree of intermittency may vary depending on solar wind conditions. Intermittency can be interpreted as a manifestation of the multifractal nature of the turbulent cascade.
A multifractal structure in the magnetic field strength $B$ has been observed at various heliocentric distances and across different phases of the solar cycle \citep{Burlaga1991, Burlaga2003, Burlaga2004b}. The foundational theory of multifractals has been explored extensively in the literature; see, for example, \citep{Mandelbrot1972,Anselmet1984}. The origin of multifractality in the solar wind may be attributed to the extension of intermittent turbulence to larger spatial scales at greater distances from the Sun, or it may arise from the nonlinear evolution and interaction of large-scale structures such as corotating streams, ejecta, and shocks.
Although solar wind plasma is often treated as almost incompressible, observed correlations between velocity, temperature, and density \citep{Elliott2016,Borovsky2021} have raised the question of whether similar nonlinear or multifractal structures might also be present in proton density. In fact, spectral analysis has revealed that proton density fluctuations exhibit Kolmogorov-like power-law behavior \citep{Shaikh2010,Chen2011}. More recently, small-scale fluctuations in solar wind proton density have been shown to exhibit multifractal properties \citep{Sorriso-Valvo2017}, highlighting the need for different intermittency measures to fully characterize the small-scale cascade. 

The paper is structured as follows. In Section 2, we present the theoretical background of non-extensive statistical mechanics and its relation to multifractal structures, fat-tailed distributions, and slow relaxation processes. Section 3 details the methodology used for extracting the \(q\)-triplet parameters from solar wind proton density data. The results of the 17-year data analysis and the validation of the \(q\)-triplet are presented in Section 4. Finally, Section 5 offers a discussion of the implications of our findings and outlines future directions for research.

\section{Multifractals, fat-tail distributions, and slow relaxation processes under the view of non-extensive statistics}

In this section, we summarize the key theoretical concepts that form the basis of our data analysis methodology. We first present an overview of nonextensive statistical theory. Based on these theoretical foundations, we then describe the data analysis methodology and the algorithm employed to produce the novel results, which are discussed in detail in the next section. 

The statistical theory of Boltzmann and Gibbs (BG) is grounded in the molecular chaos hypothesis, which assumes that the system exhibits ergodic motion in its microscopic phase space. In other words, the system can explore all microscopic states allowed with equal probability. In such cases, the probability distributions are Gaussian, and the observed time series exhibit fluctuations consistent with normal diffusion processes. Equilibrium dynamics corresponds to physical states characterized by uncorrelated or weakly correlated noise.

In contrast, nonequilibrium nonlinear dynamics can exhibit strong, long-range correlations. In such regimes, Gaussian statistics are inadequate to describe the observed behavior, as the underlying phenomena follow non-Gaussian statistics and violate the assumptions of the classical central limit theorem and the law of large numbers \citep{Umarov2008, Umarov2010}.
The standard Boltzmann-Gibbs (BG) statistical theory relies on two foundational assumptions: ergodicity and thermodynamic equilibrium. However, in systems where the dynamics are chaotic, exhibit sensitivity to initial conditions, possess memory effects, or involve long-range interactions, these assumptions no longer hold. As a result, the applicability of BG statistics is limited in such contexts.
Specific theoretical difficulties on these kinds of systems are related to the fact that the parts interact with many others at long distances, so it is impossible to cut the system into almost independent pieces. Therefore, there is no distinction between bulk and surface, and consequently these systems are non-additive and non-ergodic (phase-space is not occupied uniformly). As a result, a new kind of statistics is necessary.

Since the early 1990s, nonextensive statistical mechanics has been applied in a wide range of scientific fields, demonstrating remarkable versatility and yielding multiple applications \citep{wilk00, Gell-Mann2004,Tsallis2009,tsallis09a, Vignat2009}. It has proven particularly useful in the context of astrophysics \citep{plastino1993,Chavanis1998,Scarfone2008, Sahu2012, Rosa2013,Pavlos2018,Zamora2018c,Zamora2020}.
In particular, it has been found that the non-Gaussian distributions of magnetic field strength increments and other solar wind parameters are accurately described by the q-Gaussian distributions predicted by non-extensive statistical mechanics \citep{Burlaga2004, Burlaga2004a, Burlaga2005a}.

Nonextensive statistical mechanics is based on a generalized measure of entropy $S_q$ introduced in \citep{Tsallis1988}. $S_q$ is defined as: 

\begin{equation}
S_q = \frac{k_B}{q-1}[1-\int p(x)^qdx]   , 
\end{equation}

\nd where $p$ is the probability, and $q$ is called the nonextensivity parameter. For $q = 1$, the nonextensive entropy reduces to the standard BG entropy. The q-logarithm function is defined as

\begin{equation}
\ln_q(x) = \frac{x^{1 - q} - 1}{1 - q}, \quad x > 0    
\label{qlog}
\end{equation}
  
It is easy to verify that $\ln_{q=1}(x) = \ln(x)$. The q-logarithm satisfies the following property:  

\begin{equation}
\ln_q(x_A x_B) = \ln_q(x_A) + \ln_q(x_B) + (1 - q)\ln_q(x_A)\ln_q(x_B)  
\end{equation}
  
This function generalizes the natural logarithm. It follows directly that $S_q$ can be expressed as $S_q = k_B \int \ln_q( \rho ) \, dx$.
This expression resembles the Boltzmann-Gibbs entropy.

The inverse function of equation (\ref{qlog}) is defined as the q-exponential function, given by:  

\begin{equation}
e_q(x) = \left[1 + (1 - q)x\right]_+^{1/(1 - q)}.
\end{equation}

This function generalizes the standard exponential: if \( q = 1 \), then \( e_{q=1}(x) = e^x \). The notation \( [\,]_+ \) means that the function is defined so that it vanishes for negative arguments inside the brackets, that is, \( [x]_+ = \max(x, 0) \).

Within the non-extensive theory framework, three key features, namely, non-Gaussian distributions, slow relaxation processes, and multifractal structures, are interconnected through the so-called q-triplet. This concept was first introduced in \citep{Tsallis2005, Tsallis2003}, providing a unifying framework for describing complex, nonequilibrium systems such as the solar wind.

The \( q \)-triplet has proven to be a valuable tool for analyzing time series in atmospheric and space plasma environments. It has been applied to the study of solar activity using the AE and \( D_{st} \) indices \citep{Gopinath2018}, sunspot dynamics \citep{Pavlos2012}, nonlinear analysis of the solar flare index \citep{Karakatsanis2013}, magnetospheric self-organization processes \citep{Pavlos2012a}, and nonequilibrium phase transitions in solar wind plasma dynamics during calm and shock periods \citep{Pavlos2015}.

As evidenced by the bibliography cited so far, substantial progress has been made in this field, especially in recent years. However, the probability distribution functions of the solar wind parameters, turbulence, and transport of energetic particles remain open questions to this day \citep{Viall2020}.

Empirically derived non-Gaussian distributions are becoming increasingly prevalent in space physics, as the power-law nature of various suprathermal tails is combined with more classical quasi-Maxwellian cores. In fact, q-Gaussian distributions have been used in plasma sciences long before under the name kappa distributions, which were independently proposed \citep{Maksimovic1997a, Livadiotis2016, Yoon2019, Lazar2021, Louarn2021}. However, it can be shown that the two are equivalent through a suitable transformation \citep{Livadiotis2009}. Nevertheless, the Tsallis statistical framework provides a set of mathematical and conceptual tools that go far beyond a mere modification of the distribution, making its implementation highly enriching for plasma theory in atmospheric and space environments. These non-Gaussian distributions arise naturally within the framework of nonextensive statistical mechanics, which offers a robust theoretical foundation for describing and analyzing complex systems out of equilibrium. Given the strong correspondence between empirically observed non-Gaussian distributions and the predictions of nonextensive statistics, the full suite of nonextensive statistical tools becomes available to the space physics community for investigating the non-Gaussian characteristics of particle and energy distributions observed in space \citep{Livadiotis2009}. Moreover, the applicability of these methods extends beyond the solar wind. For example, Tsallis statistics have been shown to be effective in studying ionospheric plasma \citep{Chernyshov2014, Ogunsua2018} and magnetospheric dynamics \citep{Pavlos2011, Gopinath2018}.

\subsection{Quasi-stationary attractors and $q_{stat}$ parameter}

Contrary to BG statistical mechanics, where the function of energy describing a thermal equilibrium state is characterized by a Gaussian function, a correlated quasi-equilibrium physical process can be described by the following non-linear differential equation \citep{Tsallis2009}:

\begin{equation}
\frac{d(p_i Z)}{dE_i} = -\beta(p_i Z)^{q_{stat}},
\end{equation}

which solution is

\begin{equation}
p_i = \frac{e_{q_{stat}}^{-\beta E_i}}{Z},
\end{equation}

where

\begin{equation}
\beta_{q_{stat}} = \frac{1}{k_B T}, \quad Z = \sum_j e_{q_{stat}}^{-\beta_{q_{stat}} E_j}.
\end{equation}

The probability distribution function is then given by:

\begin{equation}
p(x) \propto \left[1 - (1 - q_{stat}) \beta x^2 \right]^{\frac{1}{1 - q_{stat}}},
\label{qgausseq}    
\end{equation}

for continuous variables. The above distribution function is the so-called q-Gaussian function, and corresponds to the attracting stationary solution associated with the non-linear dynamics of the system. The stationary solutions \(p(x)\) describe the probabilistic nature of the dynamics in the attractor set in the phase space. The stationary parameter, \(q_{stat}\), varies accordingly as the attractor changes.

\subsection{Relaxation processes and the $q_{rel}$ parameter}
BG statistics is associated with the exponential relaxation of macroscopic quantities to thermal equilibrium, ie., one expects an exponential decay with a relaxation time $\tau$. If \( \Delta S \) denotes the deviation of entropy from its equilibrium value \( S_0 \), then the probability of a proposed fluctuation is given by:

\begin{equation}
p \sim \exp(\Delta S / k_B)
\end{equation}

At the macroscopic level, the relaxation toward equilibrium of a dynamical observable \( O(t) \), which describes the system's evolution in phase space, can be modeled by the general form:

\begin{equation}
\frac{d\Omega}{dt} \simeq -\frac{1}{\tau} \Omega,
\end{equation}

where

\[
\Omega(t) \equiv \frac{[O(t) - O(\infty)]}{[O(0) - O(\infty)]}
\]

is a normalized measure of the deviation of \( O(t) \) from its stationary state value. Under the nonextensive generalization, the standard exponential relaxation process is replaced by a meta-equilibrium formulation governed by:

\begin{equation}
\frac{d\Omega}{dt} = -\frac{1}{\tau} \Omega^{q_{rel}},
\end{equation}

where \( q_{rel} \) characterizes the degree of non-extensivity in the relaxation process. The solution to this equation is:

\begin{equation}
\Omega(t) = e_{q_{rel}}^{-t / \tau},
\end{equation}

where \( e_q^x \) is the \( q \)-exponential function.

\subsection{Sensibility to initial conditions and the \( q_{sens} \) parameter}
In BG statistical mechanics, systems typically exhibit exponential sensitivity to initial conditions. This behavior, known as strong chaos, is characterized by exponential divergence of nearby trajectories and quantified by one or more positive Lyapunov exponents.

In contrast, nonextensive statistical mechanics is associated with \( q \)-exponential sensitivity to initial conditions, a hallmark of weak chaos. This regime is described by a \( q \)-exponential growth governed by the nonextensivity parameter \( q_{sens} \).

The entropy production process is intimately connected with the structure of the system's attractor in phase space. This structure can be characterized by its multifractality and by the sensitivity to initial conditions, which can be modeled by the following differential equation:

\begin{equation}
\frac{d\xi}{dt} = \lambda_1 \xi + (\lambda_q - \lambda_1) \xi^{q_{sens}},
\label{senseq}
\end{equation}

where \( \xi(t) \) quantifies the divergence between nearby trajectories and \( \lambda_1 \) is the largest Lyapunov exponent. For \( \lambda_1 > 0 \) (\( \lambda_1 < 0 \)), the system is strongly chaotic (regular), while for \( \lambda_1 = 0 \) it is at the edge of chaos. \( \xi(t) \) is defined through:

\begin{equation}
\xi \equiv \lim_{\Delta x(0) \to 0} \frac{\Delta x(t)}{\Delta x(0)},
\end{equation}

with \( \Delta x(t) \) representing the distance between neighboring trajectories in phase space \citep{Tsallis2002}.

The solution to Eq. (\ref{senseq}) is given by:

\begin{equation}
\xi(t) = \left[1 - \frac{\lambda_q}{\lambda_1} + \frac{\lambda_q}{\lambda_1} e^{(1 - q_{sens})\lambda_1 t} \right]^{\frac{1}{1 - q_{sens}}}.
\end{equation}

This expression captures the nonlinear sensitivity of the system to initial conditions, and the parameter \( q_{sens} \) serves as a quantitative measure of the degree of deviation from standard exponential sensitivity.

According to Lyra and Tsallis \citep{Lyra1998}, the scaling properties of the most rarefied and most concentrated regions of multifractal dynamical attractors can be used to estimate the divergence \( \xi \) of nearby orbits, according to the first order approximation:

\begin{equation}
\xi = e_{q_{sens}}^{\lambda_{q} t} = \left[ 1 + (1 - q_{sens})\lambda_q t \right]^{\frac{1}{1 - q_{sens}}}.    
\end{equation}

 If the Lyapunov exponent \( \lambda_1 \neq 0 \) then \( q_{sens} = 1 \) (strongly sensitive if \( \lambda_1 > 0 \), strongly insensitive if \( \lambda_1 < 0 \)). If the Lyapunov exponent \( \lambda_1 = 0 \) (weakly sensitive) then \( q_{sens} < 1 \).

\subsection{The q-triplet}
Consider the three distinct features of nonlinear systems discussed earlier. The set \((q_{stat}, q_{rel}, q_{sens})\) constitutes what is known as the q-triplet (also occasionally referred to as the q-triangle) \citep{Gell-Mann2004}. The values of the q-triplet characterize the attractor set of the dynamics in phase space. In the case of equilibrium (i.e., Boltzmann–Gibbs statistics), the q-triplet takes the values \((q_{stat} = 1, q_{rel} = 1, q_{sens} = 1)\).

These indices are interrelated, as they all arise from the particular way in which the system explores its phase space \citep{Gazeau2019}. In the case of the solar wind, the following relationships hold:

\begin{equation}
\frac{1}{q_{rel} - 1} = \frac{1}{q_{sens} - 1} + 1
\end{equation}

\begin{equation}
\frac{1}{q_{stat} - 1} = \frac{1}{q_{sens} - 1} + 2
\end{equation}

Hence, only one of the q-triplet indices is independent. The conjectured values of the q-triplet for the solar wind, based on the analysis in \citep{Burlaga2005}, are: \(q_{stat} = \frac{7}{4}\), \(q_{rel} = 4\), and \(q_{sens} = -\frac{1}{2}\). If we define the auxiliary quantities:

\begin{equation}
a_{sens} := \frac{1}{1 - q_{sens}} = \frac{2}{3},
\end{equation}

\begin{equation}
a_{stat} := \frac{1}{q_{stat} - 1} = \frac{4}{3},
\end{equation}

\begin{equation}
a_{rel} := \frac{1}{q_{rel} - 1} = \frac{1}{3},
\end{equation}

we also verify that:

\begin{equation}
a_{rel} + a_{stat} - a_{sens} = 1.
\label{suma}
\end{equation}

The q-triplet thus leads to a striking mathematical structure. If we define \( \epsilon \equiv 1 - q \), the q-triplet becomes equivalent to the set: \(\epsilon_{stat} = -\frac{3}{4}\), \(\epsilon_{rel} = -3\), and \(\epsilon_{sens} = \frac{3}{2}\). These values satisfy the following relationships:

\begin{equation}
\epsilon_{stat} = \frac{\epsilon_{sens} + \epsilon_{rel}}{2} \quad \text{(arithmetic mean)}
\end{equation}

\begin{equation}
\epsilon_{sens} = \left(\epsilon_{stat} \, \epsilon_{rel}\right)^{1/2} \quad \text{(geometric mean)}
\end{equation}

\begin{equation}
\epsilon_{rel}^{-1} = \frac{\epsilon_{stat}^{-1} + \epsilon_{sens}^{-1}}{2} \quad \text{(harmonic mean)}
\end{equation}

The interpretation of these intriguing relationships in terms of some underlying symmetry or analogous physical principle remains an open question \citep{Gazeau2019}.

The aim of this work is to investigate and verify these relationships. To this end, we performed a systematic analysis of large-scale fluctuations in the solar wind proton density using data collected by several spacecraft located at the L1 point. Our study focuses on identifying multifractal structures, probability distributions, and relaxation processes. Subsequently, we analyze the correlations among these three phenomena. 

The novelty of this study lies in the fact that this is the first systematic investigation of the q-triplet in solar wind proton density, based on continuous data spanning 17 consecutive years. Previous works have already provided evidence supporting the q-triplet framework in astrophysical and atmospheric systems, but such studies have typically been restricted to specific years or conditions (see, e.g., \citep{Burlaga2005,Ferri2010}). Here, we interpret our results in the context of nonextensive statistical mechanics, which appears to be consistent with the observed nonlinear structure of the data.

\section{Data Analysis}
\begin{figure}[h!]
    \centering
    \includegraphics[width=.5\linewidth]{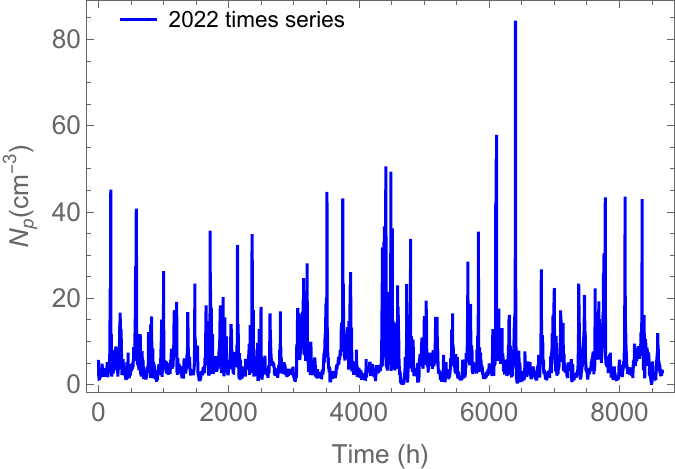}
    \caption{Time-series $N_p(t)$. Hourly averages of the proton density as a function of time, year 2022. The data was taken from the OMNI directory.}
    \label{timeserie}
\end{figure}

Let us now consider some specific observations of the fluctuations of proton density in solar wind. The data we utilize in this study was taken from the OMNI directory \citep{King2005}, https://omniweb.gsfc.nasa.gov, which contains the hourly mean values of the interplanetary magnetic field (IMF) and solar wind plasma parameters measured by various spacecraft near the Earth's  orbit. We used the low resolution data set, which is primarily a 1963-to-current compilation of hourly-averaged, near-Earth solar wind magnetic field and plasma parameter data from several spacecraft in geocentric or L1 (Lagrange point) orbits. In particular, since 2004, the priority data is taken from two spacecrafts: Wind \citep{Kasper2002} and ACE \citep{McComas1998}.
As an example, Fig. (\ref{timeserie}) shows observations of the hourly averages of proton density $N_p$ in solar wind from day 1 to 365, year 2022.

As can be seen, the fluctuations in $N_p$ are large during this interval, that is, the amplitudes of the fluctuations are larger than the mean. For each year between 2008 and 2024, we want to deduce the parameters $q_{stat}, q_{rel}$, and $q_{sens}$.

\subsection{Determination of $q_{stat}$}
The value of \( q_{stat} \) is derived from a probability distribution function (PDF). The successive fluctuations in \( N_p \) can be described by the PDFs of

\begin{equation}
dN_p(i) \equiv N_p(i + 1) - N_p(i),
\end{equation}

\nd properly normalized using the moving average \( \langle N_p(i) \rangle = \frac{N_p(i+1) + N_p(i)}{2} \). Our statistical analysis is based on the algorithm described in \citep{Ferri2010}. The range of \( dN_p \) is subdivided into small "cells" (a data binning process) of width \( \delta N_p \), in order to evaluate the frequency of \( dN_p \) values falling within each bin. The choice of bin width is a crucial step in the algorithmic process and is equivalent to solving the binning problem: a proper initialization of the bin size can significantly accelerate the statistical analysis and promote convergence of the algorithm toward the correct solution. In our case, we used the Sturges' method.

\begin{figure}[h!]
    \centering
    \begin{subfigure}[b]{0.49\textwidth}
        \centering
        \includegraphics[width=\linewidth]{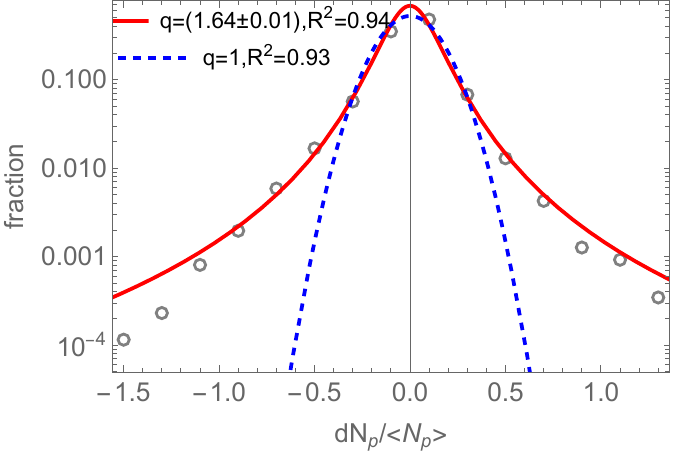}
        \caption{}
        \label{qgauss}
    \end{subfigure}
    \hfill
    \begin{subfigure}[b]{0.49\textwidth}
        \centering
        \includegraphics[width=\linewidth]{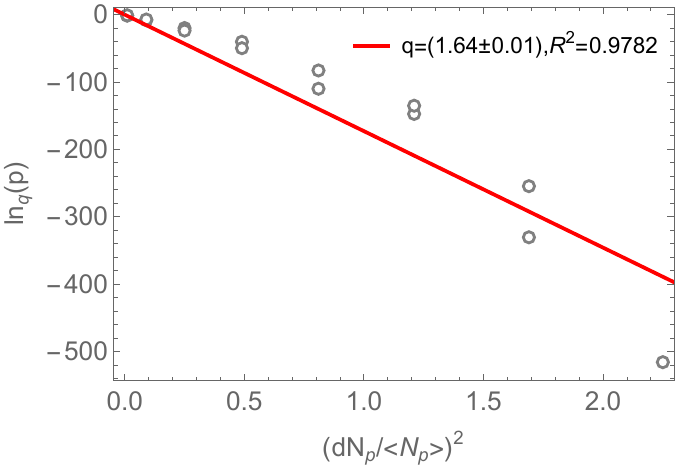}
        \caption{}
        \label{qgausslin}
    \end{subfigure}
    \caption{(a) The circles are PDFs of relative hourly changes in the proton density for the year 2022. The red solid curve is a nonlinear fit of the data with a q-Gaussian and the blue dashed curve is a Gaussian distribution. (b) Linear correlation between $ln_q(p)$ and $(dN_p/<N_p>)^2$ with $q_{stat}=1.64\pm 0.01$.}
    \label{fig1}
\end{figure}

The PDF observed for the year 2022 is shown in Fig. \ref{qgauss} as an example. The solid curve represents the best fit of the PDF to the \( q \)-Gaussian distribution (Eq. \ref{qgausseq}). The \( q \)-Gaussian distribution provides an excellent fit to all observed PDFs across the years 2008–2024.

For an initial assessment, we perform a fast nonlinear fit of the PDF using a \( q \)-Gaussian (Eq.~\ref{qgausseq}) to obtain a preliminary estimate \( q' \). Since this method typically yields an error of around 20\%, we reduce the uncertainty by linearizing the PDF. To do so, we consider the plot of \( \ln_q(p) \) versus \( (dN_p / \langle N_p \rangle)^2 \), as shown in Fig.~\ref{qgausslin}.

To refine the estimate, we vary \( q \) in steps of \( \delta q = 0.01 \) around the initial value \( q' \), performing a linear regression at each step and calculating the corresponding correlation coefficient (CC). The value of \( q \) that yields the highest CC is selected as the best estimate of \( q_{stat} \).

\subsection{Determination of $q_{rel}$}

To estimate \( q_{\text{rel}} \), one can analyze the decay of specific observables \( \Omega(t) \), such as the autocorrelation function \( C(\tau) \) or the mutual information \( I(\tau) \). The value of \( q_{rel} \) can be determined from a scale-dependent correlation coefficient \( C(\tau) \), defined as follows:

\begin{equation}
C(\tau) \equiv \frac{\langle [N_p(t_i + \tau) - \langle N_p(t_i) \rangle] \cdot [N_p(t_i) - \langle N_p(t_i) \rangle] \rangle}
{\langle [N_p(t_i) - \langle N_p(t_i) \rangle]^2 \rangle},
\end{equation}

\begin{figure}[h!]
    \centering
    \includegraphics[width=.5\linewidth]{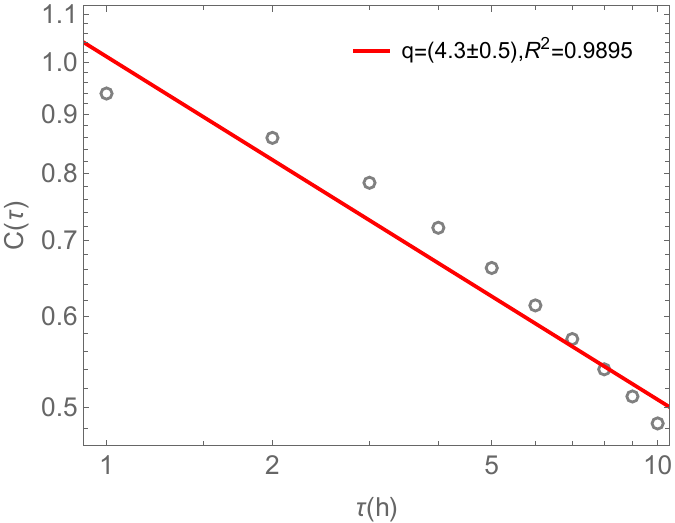}
    \caption{The autocorrelation coefficient $C(\tau)$ versus scale $\tau$ computed from hourly averages of proton density for the year 2022. The red solid line is the best fit to the data in the range 1 to 10 hours with a q-exponential, $q_{rel}=4.3\pm0.5$. \label{coefcorr}}
\end{figure}

According to non-extensive statistics, it should decay as a power law, i.e. \( \log C(\tau) = a + s \log \tau \), where the slope \( s = 1/(1 - q_{rel}) \), and \( q_{rel} \) characterize a relaxation process. 
In Fig. (\ref{coefcorr}) we show an example (year 2022), where the relaxation exhibits a power-law decay on scales from 1 to 10 hours.

\subsection{Determination of $q_{sens}$}

\( q_{sens} \) can be derived from the multifractal spectrum \( f(\alpha) \) of the attractor associated with the nonlinear dynamical system. The sensitivity to initial conditions in nonlinear systems is described by a \( q \)-exponential distribution with \( q = q_{sens} \), rather than an exponential distribution, as is the case for strong chaos.

To investigate the presence of a multifractal structure in the time series, we plot the moments of \( N_p \) at various time scales \( \tau = 2^n, n=0,1,2,3,... \). For a given value of \( \tau \), we calculate the mobile averaged value \( \langle N_p \rangle \) over the time interval \( \tau \). From this series, we construct the moments \( N_p^k \), where \( k \) is any positive or negative number. In standard multifractal analysis, the notation \( q \) is used for these moments; however, we use \( k \) here to avoid confusion with the nonextensivity parameter \( q \).

\begin{figure}[h!]
    \centering
    \includegraphics[width=.5\linewidth]{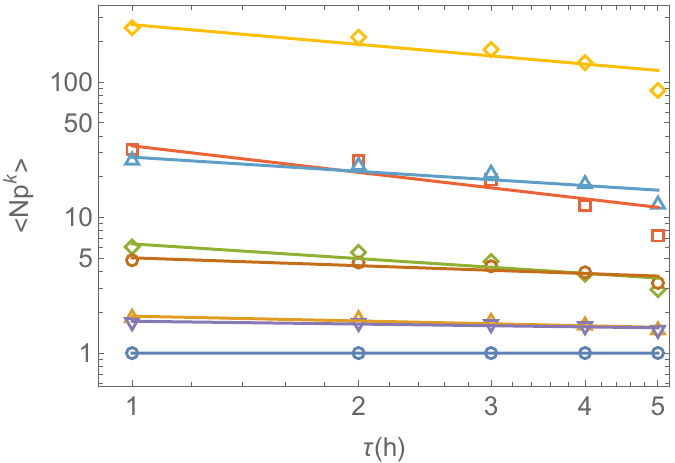}
    \caption{The \( k \)-th moments of various mobile averages of \( N_p \) as a function of scale for the year 2022. A range of scales is observed in which the points for a given moment \( k \) lie close to a straight line. From bottom to top the lines correspond to the values $k=1, k=-1, k=2, k=-2, k=3, k=-3, k=4, k=-4$. The absolute value of the slope increases with increasing \( |k| \), indicating the existence of multifractal structure.}
    \label{momentos}
\end{figure}

The result is a curve of the \( k \)-th moment of \( N_p \) as a function of scale. Finally, we repeat this procedure for multiple values of \( k \), yielding a family of curves - one for each value of \( k \) - as shown in Fig.~\ref{momentos}. These curves are straight lines on a log-log plot, and the slope increases with the magnitude of \( k \), indicating the presence of a multifractal structure over the analyzed range of scales.

This yields a set of slopes \( (k_i, s_i) \), which can be described by a nonlinear function \( s(k) \). In other words, if the proton density profile exhibits a multifractal structure, then

\begin{equation}
    \langle N_p^k \rangle \sim \tau^{s(k)}.
\end{equation}

\begin{figure}[h!]
    \centering
    \begin{subfigure}[b]{0.49\textwidth}
        \centering
        \includegraphics[width=\linewidth]{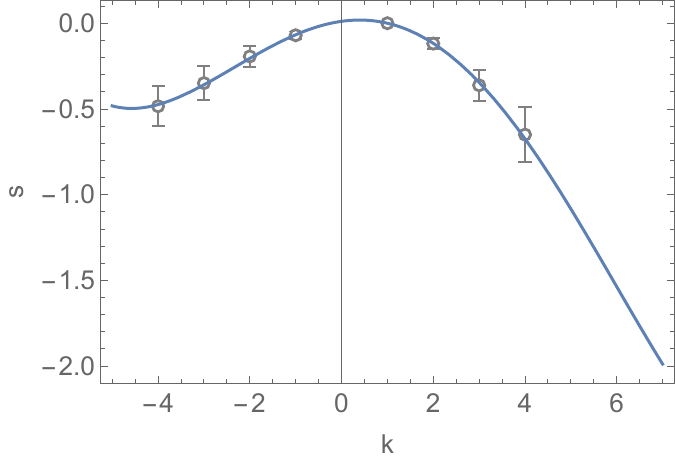}
        \caption{}
        \label{multifractal}
    \end{subfigure}
    \hfill
    \begin{subfigure}[b]{0.49\textwidth}
        \centering
        \includegraphics[width=\linewidth]{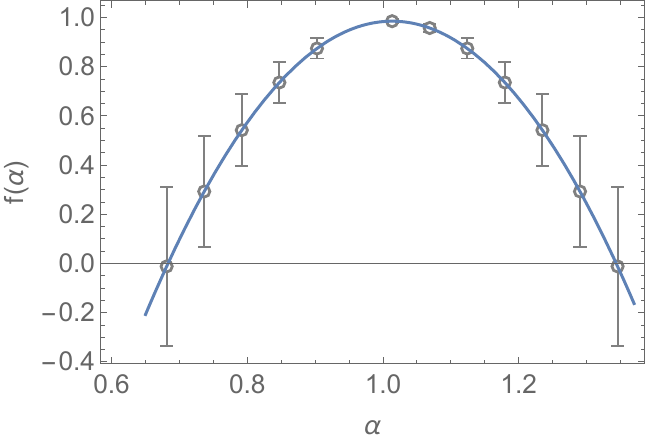}
        \caption{}
        \label{multifractal2}
    \end{subfigure}
    \caption{(a) The points \( (k_i, s_i) \) with error bars, derived from the slopes in Fig.~\ref{momentos}. The solid curve is a quadratic fit illustrating the deviation of the data. (b) The multifractal spectrum \( f(\alpha) \) derived from the same data. The solid curve is a quadratic polynomial fit used to determine the zeros $\alpha_{max}$ and $\alpha_{min}$.}
\end{figure}

The function \( s(k) \) characterizes the specific multifractal structure. The set of observed points \( (k_i, s_i) \) can be approximated well by a polynomial function \( s(k) \), so that just a few coefficients are sufficient to describe the multifractal, as shown in Fig.~\ref{multifractal}. According to \citep{Mandelbrot1972, Mandelbrot1989}, \( s(k) \) is a quadratic polynomial when the time series follows a log-normal distribution. The variance of the log-normal distribution obeys a scaling symmetry \citep{Gupta1991}. In our case, the data deviate from the quadratic fit (see the solid blue line in Fig.~\ref{multifractal}), which is expected since our distribution is not log-normal, but rather \( q \)-Gaussian. In practice, one fits the lowest-degree polynomial that provides a good fit to the data. In our example from the year 2022, we use a 4th-degree polynomial.

It is useful to introduce two additional descriptions. The first is the "generalized dimension" \( D_k(k) \) \citep{Hentschel1983}, which is related to \( s(k) \) by the equation

\begin{equation}
    D_k(k) = 1 + \frac{s(k)}{k - 1}.
    \label{dk}
\end{equation}

\( D_k \) describes the Rényi generalized dimensions, defined as

\begin{equation}
D_k = \frac{1}{k - 1} \lim_{\lambda \to 0} \frac{\log \sum_{i=1}^N p_i^k}{\log \lambda},
\end{equation}

\nd where \( p_i \) is the local probability at location \( i \) in phase space, and \( \lambda \) is the local scale. The Rényi \( k \)-indices (typically denoted \( q \), but we use \( k \) here to avoid confusion) can take values across the entire real line, \( (-\infty, +\infty) \).

The second description is given in terms of the multifractal spectrum \( f(\alpha) \) \citep{Halsey1986}, defined by the relations:

\begin{equation}
    \alpha = \frac{d}{dk} \left[ (k - 1) D_k(k) \right],
    \label{alpha}
\end{equation}

\begin{equation}
    f(\alpha) = k \alpha(k) - (k - 1) D_k(k),
    \label{efe}
\end{equation}

where \( \alpha \) is known as the Hölder exponent. Using the coefficients of the fitted polynomial \( s(k) \), we calculate a set of points in the multifractal spectrum \( f(\alpha) \) using Eqs.~(\ref{alpha}) and~(\ref{efe}). The resulting spectrum \( (\alpha, f(\alpha)) \) is shown in Fig.~\ref{multifractal2}.

The extremes of the spectrum, \( \alpha_{min} \) and \( \alpha_{max} \), for which \( f(\alpha) = 0 \), are related to \( q_{sens} \) \citep{Lyra1998, Tsallis2003} according to:

\begin{equation}
\frac{1}{1 - q_{sens}} = \frac{1}{\alpha_{min}} - \frac{1}{\alpha_{max}}.
\end{equation}

To determine \( \alpha_{min} \) and \( \alpha_{max} \), it is necessary to fit the observations with a suitable function and identify the intersection with the x axis, extrapolating \( f(\alpha) \) if necessary. The uncertainties in \( \alpha_{min} \) and \( \alpha_{max} \) propagate to the uncertainty in \( q_{sens} \), but these are largely influenced by the fitting function chosen. Although the theoretical form of \( f(\alpha) \) is not known, it is expected to be a concave function with a single maximum \citep{Beck1993}. A quadratic function, shown by the curves in Fig.~\ref{multifractal2}, provides a good fit to our observations, although the fit is not unique. A cubic fit also performs well over the observed range, but its extrapolation yields an unphysical inflection point \citep{Burlaga2005}. For the year 2022, using a quadratic fit, we obtain a value of \( q_{sens} = -0.38\pm 0.02 \).

\section{Results}

After performing the analysis described in the last section for the 17 years under consideration (2008-2024), we present in Table \ref{tabla} the results of the q-triplet for each year, and the average of all of them.

\begin{table}[h!]
\centering
\caption{Yearly values of $q_{stat}$, $q_{rel}$, and $q_{sens}$ from 2008 to 2024.}
\begin{tabular}{|c|c|c|c|}
\hline
\textbf{Year} & $q_{stat}$ & $q_{rel}$ & $q_{sens}$ \\
\hline
2008 & $1.61 \pm 0.01$ & $4.2 \pm 0.3$ & $-0.75 \pm 0.01$ \\
2009 & $1.80 \pm 0.02$ & $3.8 \pm 0.3$ & $-0.45 \pm 0.02$ \\
2010 & $1.76 \pm 0.02$ & $4.5 \pm 0.4$ & $-0.54 \pm 0.02$ \\
2011 & $1.83 \pm 0.02$ & $3.9 \pm 0.2$ & $-0.25 \pm 0.03$ \\
2012 & $1.69 \pm 0.01$ & $4.2 \pm 0.3$ & $-0.36 \pm 0.02$ \\
2013 & $1.75 \pm 0.02$ & $5.4 \pm 0.5$ & $-0.52 \pm 0.02$ \\
2014 & $1.69 \pm 0.01$ & $5.1 \pm 0.6$ & $-0.76 \pm 0.01$ \\
2015 & $1.70 \pm 0.01$ & $4.4 \pm 0.5$ & $-0.20 \pm 0.03$ \\
2016 & $1.73 \pm 0.01$ & $4.7 \pm 0.4$ & $-0.68 \pm 0.01$ \\
2017 & $1.68 \pm 0.01$ & $3.7 \pm 0.3$ & $-0.17 \pm 0.04$ \\
2018 & $1.55 \pm 0.01$ & $4.0 \pm 0.4$ & $-0.34 \pm 0.02$ \\
2019 & $1.74 \pm 0.01$ & $4.0 \pm 0.4$ & $-0.47 \pm 0.02$ \\
2020 & $1.74 \pm 0.01$ & $4.8 \pm 0.5$ & $-0.69 \pm 0.01$ \\
2021 & $1.71 \pm 0.01$ & $4.4 \pm 0.4$ & $-0.40 \pm 0.01$ \\
2022 & $1.64 \pm 0.01$ & $4.7 \pm 0.5$ & $-0.38 \pm 0.02$ \\
2023 & $1.72 \pm 0.01$ & $4.3 \pm 0.4$ & $-0.20 \pm 0.03$ \\
2024 & $1.81 \pm 0.02$ & $4.3 \pm 0.4$ & $-0.22 \pm 0.01$ \\
\hline
\textbf{Average} & $1.71\pm 0.07$  & $4.4\pm 0.5$ & $-0.4 \pm 0.2$\\
\hline
\end{tabular}
\label{tabla}
\end{table}

As we can see, the results confirm the theoretical conjectures and previous experimental findings \citep{Burlaga2005, Gazeau2019}. Note that the $q$-triplet values for each individual year do not coincide, within the error bars, with the theoretical predictions; however, the average values do. This reflects the dispersion in the values obtained, particularly in the determination of $q_{sens}$, a fact that is evident from the relatively large standard deviation.

\begin{figure}[h!]
    \centering
    \includegraphics[width=.5\linewidth]{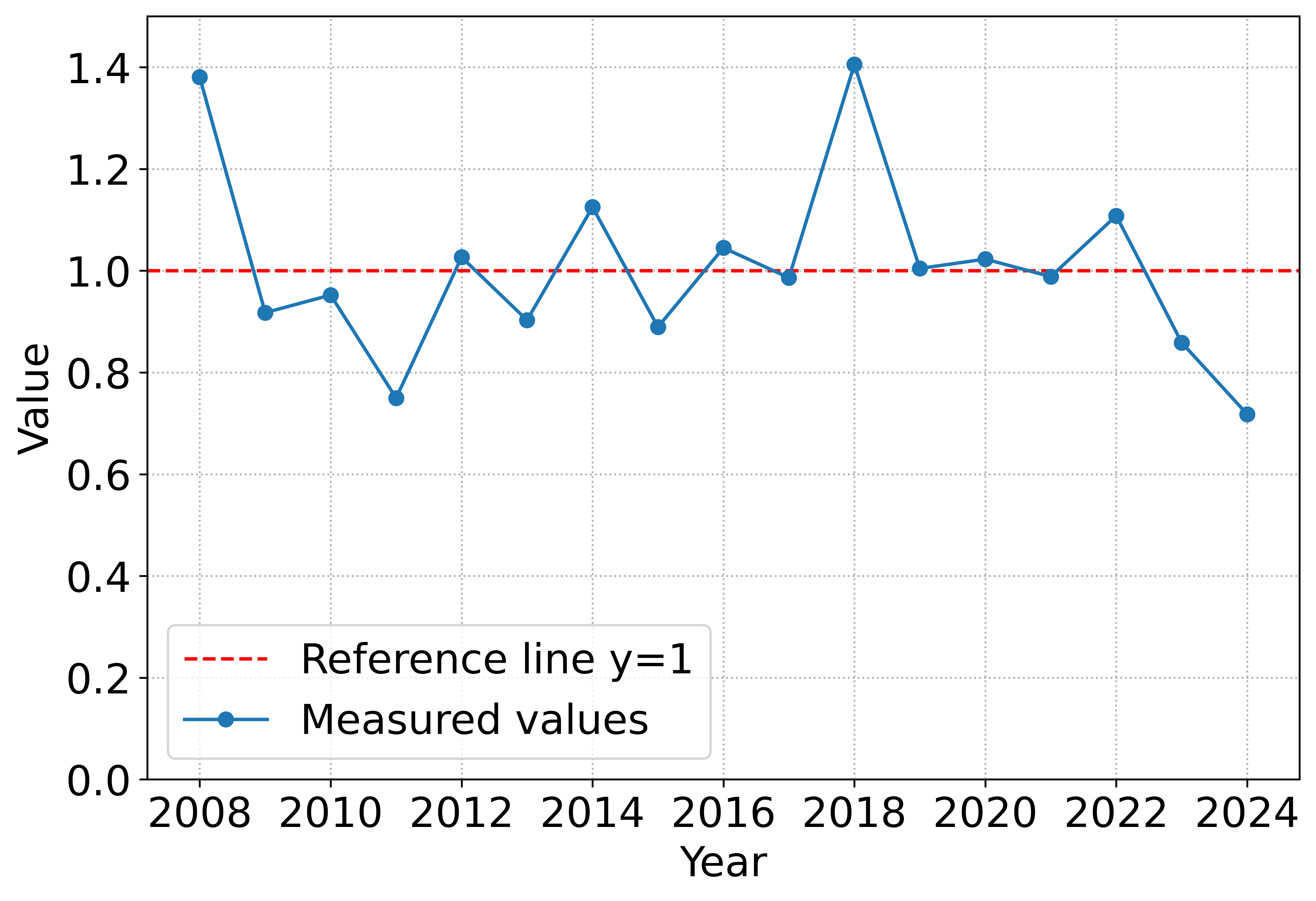}
    \caption{Value of the quantity $a_{rel}+a_{stat}-a_{sens}$ vs year. According to the theory, the q-triplet holds the relationship $a_{rel}+a_{stat}-a_{sens}=1$ for solar wind.}
    \label{sumagraph}
\end{figure}

Furthermore, in Fig. \ref{sumagraph} we show a plot of the quantity $a_{rel}+a_{stat}-a_{sens}$ as a function of the year, according to Eq. \ref{suma} should be equal to 1. The average over the range studied here is $(1.0 \pm 0.2)$.

It is important to recognize that the estimation of intermittency is subject to several uncertainties related to measurement quality, the length of the time series, and the spectral characteristics of the fluctuations, as previously noted by other authors \citep{Sorriso-Valvo2017, Viall2020}. A hypothesis regarding the variability in the values of the \( q \)-triplet is its possible dependence on the solar cycle. For example, in \citep{Pavlos2015}, the \( q \)-triplet was studied during both shock and calm periods in the solar wind, revealing different values for each regime. However, the cited work was based on high-resolution data, focusing on small-scale fluctuations, rather than investigating the long-term (large temporal scale) dependence of the \( q \)-triplet on solar activity. The study of the relationship between the \( q \)-triplet and solar activity remains an active topic of research, and we intend to present our findings on this topic in a forthcoming publication.

\section{Conclusions}

In this work, we have performed a comprehensive, year-by-year analysis of solar wind proton density fluctuations at the L1 point (near 1 AU), covering 17 consecutive years from 2008 to 2024. Using the framework of nonextensive statistical mechanics, we examined the presence and behavior of three key features of nonlinear dynamical systems: fat-tailed probability distributions, long relaxation processes, and multifractal structures. These correspond, respectively, to the indices \( q_{stat} \), \( q_{rel} \), and \( q_{sens} \) of the Tsallis \( q \)-triplet.

Our results confirm both theoretical conjectures and earlier empirical studies \citep{Burlaga2005, Gazeau2019}. Although the individual annual values of the \( q \)-triplet fluctuate and do not always match the theoretical expectations within their uncertainties, the average values over the full 17-year period do align with the predicted relationships among the indices. This agreement suggests that the Tsallis triplet structure is indeed a robust description of the solar wind’s complex behavior, and that the variability seen on a yearly basis may reflect both measurement limitations and natural dynamical fluctuations.

q-triplet have been validated against data obtained by astrophysical observations, such as those cited here, atmospherical observations (see for example, \citep{Ferri2010, Ferri2017}, and seismogenesis observations \citep{Iliopoulos2012, Pavlos2014}. A good summary of these findings is made in \citep{Pavlos2018}. All of them share in common the fact that we have no control over the variables, and therefore the measurements are noisy. Future research should aim to obtain higher-quality data and more systematic statistical analysis. These would allow for a more comprehensive comparison of different measures, leading to a deeper understanding of the nature of solar wind nonlinear character. This, in turn, would further strengthen the empirical support for the predictions of the q-triplet and other nonextensive theoretical frameworks. Another suggestion to improve the results is to search for experimental evidence in other phenomena in which variable control is possible.

In particular, we found that the standard deviation is relatively large for \( q_{sens} \), reflecting the intrinsic difficulty of estimating this index from multifractal spectra, which depend sensitively on the fitting method and extrapolation of \( f(\alpha) \). Despite this, the relationship \( q_{sens} < 1 < q_{stat} < q_{rel} \), previously noted in solar wind magnetic field studies, is also preserved in our analysis of proton density.

Our findings are especially significant in light of the differences in observational context: while previous studies were based on measurements of the interplanetary magnetic field (IMF) at heliocentric distances ranging from 7 to 87 AU, our study focuses on a plasma variable—proton density—measured continuously near Earth. The fact that the nonlinear character of the solar wind (as captured by the \( q \)-triplet) persists even at 1 AU highlights the relevance of these dynamics for understanding near-Earth space weather phenomena.

Our results suggest that long-range correlations, multifractal structure, and slow relaxation processes in the solar wind must be accounted for when modeling its interaction with the Earth's magnetosphere and the resulting space weather effects. Such nonlinear features may influence critical technologies such as satellite navigation (e.g., GPS), communication systems, and power grids.

Future work may focus on applying the same analysis to other plasma parameters (such as velocity or temperature), investigating shorter temporal windows associated with specific events (e.g., coronal mass ejections), or exploring connections between the \( q \)-triplet and geomagnetic indices. Additionally, refined statistical techniques—such as ensemble grouping by solar wind regime or solar cycle phase—may help reduce dispersion in the estimated \( q \)-values and strengthen the predictive power of this framework.

\section{Data availability}

Publicly available datasets were analyzed in this study. This data can be found here: https://omniweb.gsfc.nasa.gov. We also cited the main papers where the data is presented in the text.

\section*{Acknowledgment}
We acknowledge support from G. Ferri with the analysis code, as well as GSFC/SPDF and OMNIWeb for making the data available. This work received financial support from CONICET (Argentinian agency).

\bibliographystyle{apalike}
\bibliography{atmosfera}

\end{document}